\documentstyle[12pt]{article}
\begin{document}
\date{\today}
\begin{center}
{\bf See-saw mechanism and four light neutrino states} \\
\vspace{2 cm}
{M.Czakon, J.Gluza and M.Zra\l ek}
\vskip 1.5cm
Department of Field Theory and Particle Physics \\
Institute of Physics, University of Silesia, Uniwersytecka 4\\
PL-40-007 Katowice, Poland

\vspace{2 cm}

\newcommand{\half}{{\textstyle{1\over2}}}
\newcommand{\quar}{{\textstyle{1\over4}}}
\newcommand{\eigh}{{\textstyle{1\over 8}}}
\newcommand{\six}{{\textstyle{1\over3!}}} 
\newcommand{\Ii}{{\textstyle{1\over i}}}
\newcommand{\onestwo}{{\textstyle{1\over\sqrt{2}}}}
\newcommand{\beq}{\begin{equation}}
\newcommand{\eeq}{\end{equation}}
\newcommand{\bi}{\begin{itemize}}
\newcommand{\ei}{\end{itemize}}
\newcommand{\beqar}{\begin{eqnarray}}
\newcommand{\eeqar}{\end{eqnarray}}

\newcommand{\boldnab}{\bbox{\nabla}}
\newcommand{\boldr}{{\bf r}}

\newcommand{\al}{\alpha}
\newcommand{\bb}{\beta}
\newcommand{\g}{\gamma}
\newcommand{\del}{\delta}
\newcommand{\e}{\epsilon}
\newcommand{\mn}{{\mu\nu}}
\newcommand{\pa}{\partial}
\newcommand{\npa}{\rlap{/}\partial}
\newcommand{\nA}{\rlap{/}\!A}
\newcommand{\na}{\rlap{/}a}
\newcommand{\nk}{\rlap{/}k}
\newcommand{\np}{\rlap{/}p}
\newcommand{\tr}{\mathop{\rm tr}}
\newcommand{\dd}[1]{\mathop{{\rm d}#1}}
\newcommand{\dds}[2]{\mathop{{\rm d\null}^{#1}#2}}
\newcommand{\fract}[2]{{\textstyle{\frac{#1}{#2}}}}

\let\epsilon\varepsilon
\let\bigl\Bigl
\let\bigr\Bigr

{\bf Abstract}
\end{center}

A formal proof is given that in a see-saw type neutrino mass matrix with only two 
neutrino mass scales ($m_D \ll m_R$) and the maximal rank of $m_{R(D)}$
we can not get a fourth light sterile neutrino. 
\vspace{2 cm}





It has been known for long that extended gauge group models, such as
$SO(10)$ \cite{so10} or $SU(2)_L \times SU(2)_R \times
U(1)_{B-L}$ \cite{lr}, naturally develop a see-saw type neutrino mass matrix. 
Namely ($m_D$ is a $3\times n_R$ matrix and $m_R$ is a $n_R \times n_R$ matrix)
\begin{equation}
m_\nu = \left( \matrix{ 0 & m_D \cr 
m_D^T & m_R } \right),
\label{first}
\end{equation}
coupled with a large scale difference between $m_D$ and $m_R$, yields a mass
spectrum containing three light Majorana neutrinos, effectively described by
\begin{equation}
m_{light} \simeq m_D^T m_R^{-1} m_D.
\end{equation}
This leads to two different light $\Delta m^2$ mass scales, enabling one to understand
the solar and the atmospheric neutrino anomalies 

According to the current data (now at 99 \% c.l. \cite{99}) a fourth sterile 
neutrino is not necessary to explain the Superkamiokande data.
Suppose however that the LSND experiment \cite{lsnd} is confirmed and we wish also to explain 
its data in a language of neutrino oscillation phenomena. 
We then have to introduce a fourth light neutrino of sterile nature
(due to the invisible width
measurement at LEP). 
To avoid fine tunings of parameters and still have another
light neutrino, one requires the theory to have additional properties, like 
approximate horizontal symmetry \cite{smi},
exact parity symmetry \cite{par}, a discrete $Z_5$ symmetry \cite{z5}, global 
$S_3 \times Z_2$ symmetry \cite{gibb}, or even additional gauge
 ($SU(2)_S$) symmetry \cite{add}. For more examples see \cite{rev}.
Some phenomenological considerations have also appeared (see e.g. \cite{phen}).
Interestingly enough, a see-saw type mass matrix Eq.~(\ref{first}) 
can also lead to the fourth light neutrino.
This is realized by the so-called   singular see-saw mechanism \cite{sing,bi}.
The goal is achieved, by having an $m_R$ of rank $n_R-1$. However, this is not
enough, we still have to fine tune $m_R$ to the keV-MeV range. This last
unwanted problem can be circumvented by building a second stage of see-saw
structure. This still fits into the scheme Eq.~\ref{first}, but there are
in fact three scales, not two. 

Here we give a formal proof that with only two scales we can not get a fourth light
neutrino. The importance of this result lies in the fact that one may be tempted
to believe that the larger the mass matrix the more possibilities of choosing the masses
are available, and some symmetries may help getting light sterile neutrinos. 
The problem is defined in the following statement:



Let $m_R$ be a matrix
of eigenvalues real positive and greater than some $M$, and let all of the moduli of elements 
of $m_D$ be much smaller than $M$, then the
spectrum of $m_\nu$ contains the full spectrum of $m_R$ with corrections of the order of
$m^2/M\equiv max(|(m_D)_{ij}|)^2/M$. That means that no
manipulation on $m_D$ and/or $m_R$ 
can move a mass from the heavy $m_R$ matrix into the light spectrum. 

The proof is a simple consequence of perturbation theory for finite
matrices \cite{reedsimon}. There, it is shown, that if we have a matrix of
the form 
\begin{equation}
M(\beta) = M^{(0)}+\beta M^{(1)},
\end{equation}
then, every non-degenerate eigenvalue of $M^{(0)}$ gives rise to a non-degenerate
eigenvalue of $M(\beta)$, being an analytic function of $\beta$ in some surrounding of zero.
Since we are interested in the heavy spectrum, the assumption of non-degeneracy
is quite general. In case of degenerate eigenvalues, we still can expand
the eigenvalues in series, which however will be analytic functions with branches.
We limit ourselves to the former case, but the reader should understand
that the theorem holds for the general case also.

We decompose $m_\nu$ as
\begin{eqnarray}
\left( \matrix{ 0 & m_D \cr
 m_D^T & m_R
} \right) &=& \left( \matrix{ 0 & 0 \cr
 0 & m_R
} \right)+ \left( \matrix{ 0 & m_D \cr
 m_D^T & 0
} \right) \nonumber \\ \nonumber \\
&=& M \left[ \left( \matrix{ 0 & 0 \cr
 0 & m_R/M
} \right)+ \beta \left( \matrix{ 0 & m_D/m \cr
 m_D^T/m & 0
} \right) \right]\nonumber \\ \nonumber \\
&\equiv & M(M^{(0)}+\beta M^{(1)}),
\end{eqnarray}
where $\beta =m/M$. The first matrix has all its elements greater than one, while
the second has all elements smaller than one, both are of course dimensionless.
The eigenvectors of $M^{(0)}$ are of the form (we chose without loss of generality 
a weak base for neutrinos in which $m_R$ is diagonal \cite{gl})
\begin{equation}
e_i = \left( \matrix{ 0 \cr
 \vdots \cr
 1 \cr
 \vdots \cr
 0 } \right).
\end{equation}
To find the first order correction, we expand the eigenvectors as
\begin{equation}
v_i = \sum_{j} \alpha_{ji} e_j.
\end{equation}
This gives us the following equation
\begin{equation}
(M^{(0)}+\beta M^{(1)}) v_i = \lambda_i v_i,
\end{equation}
which is solved into
\begin{equation}
\lambda_i = \lambda_i^{(0)}+\beta \frac{1}{\alpha_{ii}} \sum_j \alpha_{ji} e_i^T M^{(1)} e_j,
\end{equation}
where $\lambda^{(0)}$ are non-degenerate nonzero eigenvalues of $M^{(0)}$, which are
also eigenvalues of $m_R/M$. 
Obviously, only $\alpha_{ii}$ is ${\cal O}(1)$, and
$\alpha_{ji}$ for $j\neq i$ is ${\cal O} (\beta)$. 
The first order correction to 
$\lambda^{(0)}$ is therefore
\begin{equation}
\lambda_i^{(1)} = \lambda_i^{(0)}+\beta e_i^T M^{(1)} e_i.
\end{equation}
But this vanishes due to the nondiagonal form of $M^{(1)}$. Thus the first non-vanishing
correction to the large eigenvalues is of the order $M\beta^2 = m^2/M$, which
completes the proof. Remark, that the masses of the neutrinos are moduli of the
eigenvalues of $m_\nu$. The corrections to the moduli are however of the same order.

Using simple arguments based on perturbation series, we have shown
that a natural $m_R$ (no fine tunings and eigenvalues at the heavy 
scale) leads to three light neutrinos. Thus from the class of see-saw type models 
only a singular double
see-saw mechanism can accommodate the LSND data and an additional fourth light neutrino state.

\begin{center}
{\large \bf Acknowledgments}
\end{center}
This work was supported by the Polish Committee for Scientific Research under 
Grant No. 2P03B04919  and 2P03B05418.

\end{document}